\documentclass[12pt]{article}

\usepackage[margin=1in]{geometry}
\usepackage{ctable,microtype,natbib,amsmath,amssymb,fullpage,graphicx}
\usepackage{setspace}
\usepackage{amsmath}
\usepackage{color, colortbl}
\usepackage{listings}
\usepackage{hyperref}

\date{March 12, 2016}

\definecolor{mygray}{gray}{0.6}

\title{Probabilistic Record Linkage and Deduplication after Indexing, Blocking, and Filtering}
\author{Jared S. Murray\footnote{Visiting Assistant Professor, Carnegie Mellon University Department of Statistics. Research reported in this work was supported by the National Science Foundation under grant numbers SES-1130706 and DMS-1043903. Any opinions, findings, and conclusions or recommendations expressed in this
material are those of the author(s) and do not necessarily reflect the views of the funding agencies.}}

\newcommand{\tgamma}{\gamma}
\newcommand{\tGamma}{\Gamma}
\newcommand{\tpi}{\pi}
\newcommand{\tp}{p}
 \DeclareMathOperator*{\argmax}{arg\,max}

\newcommand{\g}{g}
\newcommand{\ind}[1]{1(#1)}

\definecolor{highlighted}{RGB}{173, 235, 152}

\begin{document}

\maketitle{}
\pagenumbering{arabic}

\begin{abstract}

Probabilistic record linkage, the task of merging two or more databases in the absence of a unique identifier, is a perennial and  challenging problem. It is closely related to the problem of deduplicating a single database, which can be cast as linking a single database against itself. In both cases the number of possible links grows rapidly in the size of the databases under consideration, and in most applications it is necessary to first reduce the number of record pairs that will be compared.  

Spurred by practical considerations, a range of methods have been developed for this task. These methods go under a variety of names, including indexing and blocking, and have seen significant development. However, methods for inferring linkage structure that account for indexing, blocking, and additional filtering steps have not seen commensurate development. In this paper we review the implications of indexing, blocking and filtering within the popular Fellegi-Sunter framework, and propose a new model to account for particular forms of indexing and filtering.

\end{abstract}

\section{Introduction}

Probabilistic record linkage is the process of merging two or more databases which lack unique identifiers. The related task of detecting duplicate records in a single file can be cast as linking a file against itself, ignoring redundant comparisons. Initially developed by \cite{Newcombe1959,newcombe1962record}, probabilistic record linkage was mathematically formalized by \cite{FS}. In the ensuing decades these methods have been widely deployed, and variations on the Fellegi-Sunter framework still form the backbone of most applications of probabilistic record linkage. 

Naively matching a file with $N_A$ records to a file with $N_B$ records requires making $N_AN_B$ comparisons as an initial step; deduplicating a single file with $N$ records requires making $N(N-1)/2$ comparisons. Even if the comparisons themselves are relatively inexpensive to compute and the files are of moderate size, this step can be computationally prohibitive. In most practical applications of probabilistic record linkage it is necessary to eliminate a large number of record pairs from consideration, without making a full comparison of the two records, a process known as {\em indexing} or {\em blocking}. Storage and other considerations often lead to an additional {\em filtering} step, where record pairs that are extremely unlikely to be true matches are discarded after a complete or (nearly complete) comparison has been made.

As the size of the files under consideration has increased, the development of strategies for indexing, blocking, and filtering has outstripped the capacity of models to account for them. Using the popular Fellegi-Sunter framework as a guide, this paper discusses the implications of these strategies on subsequent modeling and inference of linkage structure. We propose extensions that provide more relevant and accurate error estimates. 

This paper proceeds as follows: Section \ref{sec:prl} reviews the mathematical formulation of probabilistic record linkage and the \cite{FS} framework, and describes some common strategies for reducing the number of record pairs. Section \ref{sec:prl-indexing} discusses the modeling and inferential implications of these strategies. Section \ref{sec:modeling} develops an extension of the \cite{FS} framework to account for the effects of some indexing methods. Section \ref{sec:example} provides illustrations on synthetic data. Section \ref{sec:conclusion} concludes with discussion about extensions and the implications for other probabilistic record linkage methods.

\section{Background: Probabilistic Record Linkage and Deduplication}\label{sec:prl}

The basic framework for linking two files is as follows: Let $A$ and $B$ be two databases, and let $a$ and $b$ generically index records in $A$ and $B$. Let $a\sim b$ denote that records $a$ and $b$ truly correspond to the same entity, and define $M=\{ (a,b)\in A\times B : a\sim b)$ and $U=\{ (a,b)\in A\times B : a\not\sim b)$. The goal is to correctly classify each record pair as a match or non-match in the absence of unique identifiers. Deduplicating a single database is similar: We consider record pairs $(a, a')$ from a single database $A$, with the goal of classifying each pair into matching and non-matching sets. In the remainder of the paper we use probabilistic record linkage to refer to linking two files as well as deduplicating a single file. 

\subsection{The Fellegi-Sunter Framework}\label{sec:fs}

The original method for probabilistic record linkage, which is still widely in use, was introduced by \cite{FS} who formalized earlier developments by \citep{Newcombe1959,newcombe1962record}. See \cite{HerzogScheurenWinkler200705} for extensive review of the basic framework and extensions.

A set of fields are available in both files $A$ and $B$ and may be used to compare records. Often these comparisons take the form of a series of binary variables, which may indicate direct matches on fields (do records $a$ and $b$ agree on gender?), sufficient agreement (is the similarity score between the two name fields greater than some threshold?) or other derived comparisons (do $a$ and $b$ match on month and year of birth?). Let $\tgamma_{ab} = (\tgamma_{ab}(1),\dots \tgamma_{ab}(q))$ be a binary vector collecting the comparisons between records $a$ and $b$, taking values in $\tGamma = \{0,1\}^{q}$. The model for record linkage presented in \cite{FS} is as follows:
\begin{gather}
\Pr[(a,b)\in M] = \tp_M\label{eq:fs-full0}\\
%
\Pr[\tgamma_{ab}=\g\mid (a,b)\in M] = \tpi_{\g\mid M}\\ 
\Pr[\tgamma_{ab}=\g\mid (a,b)\in U] = \tpi_{\g\mid U}\\
\Pr[\tgamma_{ab}=\g] = \tp_M\pi_{\g\mid M} + (1-\tp_M)\tpi_{\g\mid U}.\label{eq:fs-full1}
\end{gather}
The probability distribution of the observed comparison vectors is a two component mixture model, where one component corresponds to true matches and the other to true non-matches. The components $\tpi_{\g\mid M}$ and $\tpi_{\g\mid U}$ are often referred to as ``$m-$probabilities'' and ``$u-$probabilities'' respectively \citep{winkler2006overview}.

The parameters are usually estimated via EM \citep{winkler1988using}. The saturated model above is typically not estimable, and it is common to assume conditional independence between comparisons so that
\begin{equation}
\pi_{g\mid M} = \prod_{j=1}^p \rho_{j\mid M}^{g(j)}(1-\rho_{j\mid M})^{1-g(j)},\quad
\pi_{g\mid U} = \prod_{j=1}^p \rho_{j\mid U}^{g(j)}(1-\rho_{j\mid U})^{1-g(j)}.\label{eq:ci}
\end{equation}
Log-linear models can be used to model conditional dependence between comparisons (see e.g. \cite{thibaudeau1993discrimination}). \cite{winkler1993improved} imposed additional constraints on various probabilities to improve parameter estimation.

After estimation the parameters are used to determine the linkage structure. Each record pair is classified as a match ($A_1$), a nonmatch ($A_3$), or indeterminate ($A_2$). Indeterminate pairs are sent out for clerical review. \cite{FS} provide a decision rule that controls the following error rates: 
\begin{align}
\mu &= P(A_1\mid (a,b)\in U) = \sum_{g\in\Gamma} P(A_1\mid  \gamma_{ab}= g)\Pr[\gamma_{ab}=g \mid (a,b)\in U]\\
\lambda &= P(A_3\mid (a,b)\in M) = \sum_{g\in\Gamma} P(A_3\mid  \gamma_{ab}= g)\Pr[\gamma_{ab}=g \mid (a,b)\in M],
\end{align} 
while minimizing the number of record pairs assigned to $A_2$.
The decision rule is based on the weights
\begin{equation}
w_{ab} = \frac{\pi_{\gamma_{ab}\mid M}}{\pi_{\gamma_{ab}\mid U}},\label{eq:weights}
\end{equation}
the likelihood in favor of $(a,b)\in M$. The decision rule declares $(a,b)$ a match if $w_{ab}\geq T_\mu$, a non-match if $w_{ab} \leq\ T_\lambda$ and indeterminate if $T_\lambda<w_{ab}<T_\mu$. The two thresholds $T_\lambda$ and $T_\mu$ are set based on specified values for $\mu$ and $\lambda$. In \cite{FS} the thresholds $T_\lambda$ and $T_\mu$ are determined as follows: The set of possible comparison vectors $\gamma\in \Gamma$ is ordered such that $w_\gamma = \pi_{\gamma\mid M}/\pi_{\gamma\mid U}$ is monotonically decreasing. Index this ordered set of comparisons by $i,\ 1\leq i\leq N_\Gamma$. Then find $1\leq n\leq n'-1\leq N_\Gamma$ such that

\begin{align}
\sum_{i=1}^{n-1} \pi_{\gamma_i\mid U} < &\mu \leq \sum_{i=1}^{n} \pi_{\gamma_i\mid U}\\
\sum_{i=n'}^{N_\Gamma} \pi_{\gamma_i\mid M} \geq &\lambda > \sum_{i=n'+1}^{N_\Gamma} \pi_{\gamma_i\mid M}.
\end{align}
Assume for simplicity that there exist $n$ and $n'$ such that $\mu = \sum_{i=1}^{n} \pi_{\gamma_i\mid U}$ and $\lambda = \sum_{i=n'}^{N_\Gamma} \pi_{\gamma_i\mid M}$ (otherwise a randomized decision rule is needed, see \cite{FS} for details).  Then the thresholds are given by 

\begin{align}
T_\mu = \frac{\pi_{\gamma_n\mid M}}{\pi_{\gamma_n\mid U}},\quad
T_\lambda = \frac{\pi_{\gamma_{n'}\mid M}}{\pi_{\gamma_{n'}\mid U}}.
\end{align}

\subsection{Reducing the Number of Comparisons}

Naively matching two files requires comparing each pair of records, which is infeasible for large files even when the comparisons are computationally inexpensive. {\em Indexing} techniques quickly filter out dissimilar record pairs that are extremely unlikely to be matches \cite[Chapter~2]{Christen201207}. Two common indexing techniques are:

\begin{itemize}
\item {\bf Blocking}, which partitions records based on the values of a key like a postal code or the first initial of the last name. Blocking keys may be constructed by conjunctions of multiple keys (e.g., agreement on last initial {\em and} postal code). Record pairs are discarded unless they agree on the blocking key.
\item {\bf Indexing by disjunctions}, which retains record pairs that match on {\em one or more keys} (their disjunction). For example, we could retain only those pairs which agree on either last initial {\em or} postal code. More complex indexing schemes can be constructed using disjunctions of conjunctions. Indexing by disjunctions is typically carried out by doing multiple blocking passes using different keys and taking the union of all the retained pairs.\footnote{The terminology is not standardized in the literature; it is common for authors to ignore the distinction between what we call indexing and blocking. We follow Christen's usage here, as the distinction becomes important later.}
\end{itemize}

It is also common to discard pairs that are not excluded by indexing, but which are unlikely to be a match. So we can add to the above list

\begin{itemize}
\item {\bf Filtering}, which discards any pairs not excluded by initial indexing steps but which are still unlikely to be a match. For example, in the case of binary comparisons we might discard any pairs $(a,b)$ with $\gamma_{ab}=(0,0,\dots,0)$ or $\sum_{j=1}^p \gamma_{ab}(j) \leq 1$. Filtering rules can be more complex: For example, the U.S. Census Bureau's BigMatch software filters record pairs using initial values of the $m$ and $u$ probabilities, $\tilde\pi_{g\mid M}$ and $\tilde\pi_{g\mid U}$, and a user provided cutoff $c_0$ (dropping any pairs with $\log(\tilde\pi_{g\mid M}) -\log(\tilde\pi_{g\mid U}) < c_0$) \citep{censusyancey02}.
\end{itemize}

Filtering can be interpreted as indexing by a particular collection of disjunctions, but unlike most indexing schemes it will generally require actually performing all or nearly all of the comparisons, negating many of the computational benefits of indexing. In Section \ref{sec:example} we will see that filtering can still have significant statistical value. This is especially true in the absence of high-quality keys for indexing, or in the presence of model misspecification. 

\begin{figure}
 \centering
 \includegraphics[width=.4\textwidth]{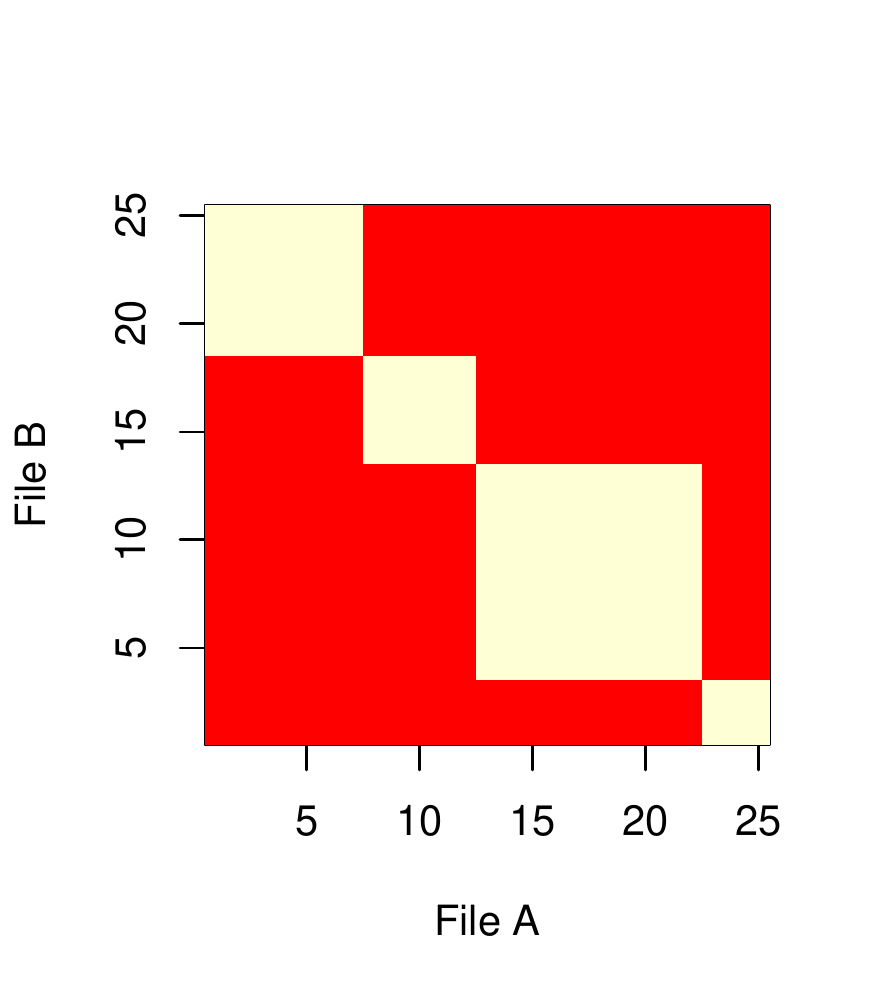}
 \includegraphics[width=.4\textwidth]{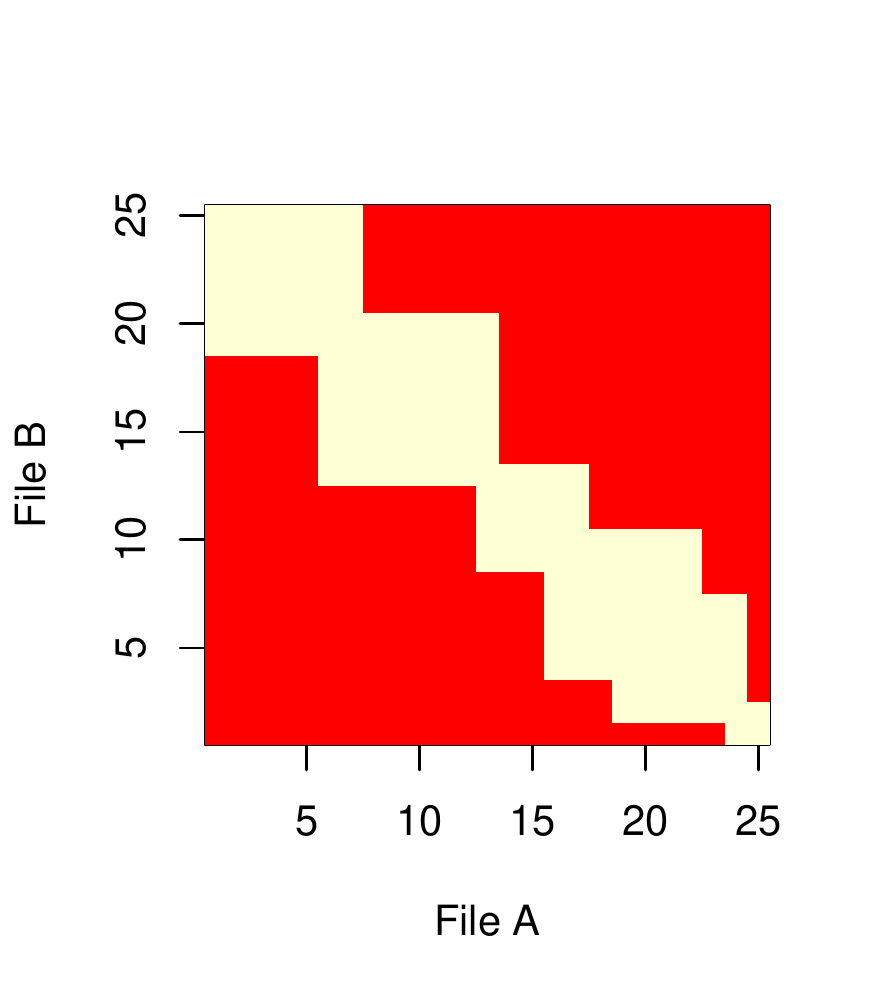}
 \caption{Comparison between blocking (left) and indexing by disjunctions (right). Record pairs in red are excluded by indexing. Blocking always partitions the {\em records} in each file so that records from one partition in file A are only allowed to match records from a corresponding partition in file B.}
 \label{fig:indexing} 
\end{figure}

Figure \ref{fig:indexing} compares blocking and indexing by disjunctions (or filtering) graphically. Observe that blocking yields a partition of records such that all links occur within and not between elements of the partition (the ``blocks'') (Fig. \ref{fig:indexing}, left). Other indexing schemes, including indexing by disjunctions, yield ``overlapping partitions'' (Fig. \ref{fig:indexing}, right). 

Indexing by disjunctions is a way to utilize multiple keys while hedging against typographical or measurement errors that would exclude true matches. Consider the records in Table \ref{tab:blocking}. Blocking on first initial of the last name captures the  ``Heather-Heather'' pair, a likely match,  but misses the ``Jane-Jane'' pair which is also a likely match. Blocking on the zip code captures the ``Jane-Jane'' pair but excludes the ``Heather-Heather'' pair. Either scheme probably introduces an error.  But indexing by the disjunction (keeping record pairs that match on zip code {\em or} first initial of last name) captures both pairs while excluding the unlikely ``Jane-Heather'' pair and both unlikely ``Paul'' pairs.

Indexing by disjunctions is also computationally efficient since it can be implemented by merging the results of multiple blocking queries. For these reasons it is widely used in practice. For example, \cite{winkler2010fast} reports on various disjunctions used in deduplicating decennial Census records and interstate voter registration rolls, and \cite{sadosky2015blocking} considered indexing by disjunctions for linking records of civilian casualties in the Syria conflict. The BigMatch software developed by the U.S. Census Bureau \citep{censusyancey02} was designed specifically to efficiently index by disjunctions, and also includes a subsequent filtering step.

More recent research has focused on more sophisticated methods to rapidly compute approximate dissimilarities between records using hash functions, or to infer blocking schemes using labeled matching/non-matching pairs. See e.g. \cite{steorts2014comparison,christen2012indexing} and \cite{baxter2003comparison} for reviews. We reserve discussion of these for Section \ref{sec:conclusion}. 

\begin{table}[]
\centering
\begin{tabular}{lllllllll}
\multicolumn{4}{c}{File A}               &  & \multicolumn{4}{c}{File B}              \\
First & Last    & Street         & Zip   &  & First & Last  & Street          & Zip   \\
Jane  & Calder   & 123 Main St    & 15210 &  & Jane  & Kalder & 123 Main Street & 15210 \\
Paul  & Frankes & 5 Birch Blvd & 15232 &  & Heather & Porter & 12 Maple Ave  & 15236 \\
Heather & Porter   & 12 Maple Ave & 51236 &  & …     & …     & …               & …     \\
…     & …       & …              & …     &  &       &       &                 &      
\end{tabular}
\caption{Example records from two files}
\vspace{0.5em}
\label{tab:blocking}
\end{table}

\section{Probabilistic Record Linkage after Indexing}\label{sec:prl-indexing}

Whether implemented by blocking, filtering or some other processs, indexing creates a biased sample of record pairs by design. For model-based procedures (including \cite{FS}) indexing therefore changes the interpretation of the recovered parameters. Researchers have long noted this fact (beginning at least with \cite{FS} themselves; related comments appear in \cite{Jaro1989} and \cite{winkler2006overview}). 
However, the effect of indexing on subsequent modeling of record pairs and inference of linkage structure is often ignored. Using the Fellegi-Sunter framework as a guide we describe some of the implications in a simple special case, comparing the effects of traditional blocking and indexing by disjunctions (including filtering). 


Let $\beta_{ab}$ be an additional binary comparison, indicating whether $(a,b)$ match on some {\em blocking} criterion. Similarly, let $\iota_{ab}$ be an additional binary comparison indicating agreement on some disjunction of other comparisons. We have in mind an initial blocking step that drops any record pairs with $\beta_{ab}\neq 1$, and subsequent indexing by disjunctions/filtering that drops additional record pairs with $\iota_{ab}\neq 1$. 

The distinction between blocking and indexing by disjunctions or filtering is important. It would be redundant to include the comparison $\beta_{ab}$ in $\gamma_{ab}$, since it is always one in the retained pairs. But when indexing by disjunctions or filtering, the comparisons comprising the disjunction { should} appear in $\gamma_{ab}$. For example, if we index by the disjunction of agreement on age and postal code, $\iota_{ab}=1$ doesn't indicate whether there was agreement on age, postal code or both. Naturally the same argument applies when filtering. We discuss the modeling implications of blocking and of indexing by disjunctions/filtering before discussing the effect of each on the estimation of error rates and decision rules.

\subsection{Modeling after Blocking}\label{sec:blockingmodel}
Under blocking those pairs with $\beta_{ab}\neq 1$ are treated as sure non-matches and are not used in parameter estimation.  This shifts our focus from $P(\gamma_{ab})$ to $P(\gamma_{ab} \mid \beta_{ab} = 1)$. Structurally the model remains identical to \eqref{eq:fs-full0}-\eqref{eq:fs-full1}: 
\begin{gather}
\Pr[(a,b)\in M\mid \beta_{ab} = 1] = p_{M\mid\beta}\label{eq:fs-block0}\\
%
\Pr[\gamma_{ab}=\g\mid (a,b)\in M, \beta_{ab} = 1] = \pi_{\g\mid M, \beta}\\ 
\Pr[\gamma_{ab}=\g\mid (a,b)\in U, \beta_{ab} = 1] = \pi_{\g\mid U, \beta}\\
\Pr[\gamma_{ab}=\g \mid \beta_{ab} = 1] = p_{M\mid\beta}\pi_{\g\mid M, \beta} + (1-p_{M\mid\beta})\pi_{\g\mid U, \beta}\label{eq:fs-block1}
\end{gather}
The parameters, however, are not the same. In particular, we expect that:
\begin{itemize}
\item $p_{M\mid\beta} >> \tp_M$, provided that blocking was effective.
\item $\pi_{\g\mid M, \beta}\approx \pi_{\g\mid M}$, since effective blocking retains (nearly) all the record pairs in $M$. 
\item $\pi_{\g\mid U, \beta}$ may be slightly smaller than $\pi_{\g\mid U}$ for comparison vectors $g$ with few ones, with a commensurate increase in the conditional probability of comparison vectors with more ones. This is due to conditioning on $\beta=1$: Given that they match on the blocking comparison, the set of retained pairs are likely to be more similar than two pairs selected at random, even if they are truly non-matches. See e.g. \cite{Jaro1989} for further discussion. \cite{Jaro1989} suggested estimating the $u-$probabilities using all pairs (or a randomly selected subset of all pairs), including those excluded by blocking, to mitigate bias indicated in the final bullet. This does not seem to be current practice, however. An alternative approach, which we pursue in this paper, is to explicitly acknowledge and account for the fact that inference is only valid for the subset of record pairs under consideration.
\end{itemize}

\subsection{Modeling after Indexing by Disjunctions/Filtering} 

When blocking is followed by another indexing step the target shifts from $P(\gamma_{ab} \mid \beta_{ab} = 1)$ to $P(\gamma_{ab} \mid \beta_{ab} = 1, \iota_{ab}=1)$ and our model becomes

\begin{gather}
\Pr[(a,b)\in M\mid \beta_{ab} = 1, \iota_{ab}=1] = p_{M\mid\beta,\iota}\label{eq:fs-index0}\\
%
\Pr[\gamma_{ab}=\g\mid (a,b)\in M, \beta_{ab} = 1, \iota_{ab}=1] = \pi_{\g\mid M, \beta, \iota}\\ 
\Pr[\gamma_{ab}=\g\mid (a,b)\in U, \beta_{ab} = 1, \iota_{ab}=1] = \pi_{\g\mid U, \beta, \iota}\\
\Pr[\gamma_{ab}=\g \mid \beta_{ab} = 1, \iota_{ab}=1] = p_{M\mid\beta,\iota}\pi_{\g\mid M, \beta,\iota} + (1-p_{M\mid\beta,\iota})\pi_{\g\mid U, \beta,\iota},\label{eq:fs-index1}
\end{gather}
which has the same structure as \eqref{eq:fs-full0}-\eqref{eq:fs-full1} and \eqref{eq:fs-block0}-\eqref{eq:fs-block1}.

The effects of indexing by disjunction or filtering are similar to those under blocking but can be more extreme. We expect that $p_{M\mid\beta, \iota} > p_{M\mid\beta} >> \tp_M$ when indexing or filtering and blocking are all effective. We also expect that $\pi_{\g\mid M, \beta, \iota}\approx \pi_{\g\mid M, \beta}\approx \pi_{\g\mid M}$, since (nearly) all of the truly matching pairs are retained. But unlike simple applications of blocking, when indexing by disjunctions or filtering the comparison space itself can change: For example, if we index by the disjunction of exact matches on age and postal code then
\[
\Pr[(a,b)\text{ disagree on age and postal code}\mid \beta_{ab}=1, \iota_{ab}=1]=0,
\]
so the {\em support} of $\gamma_{ab}$ changes when conditioning on $\iota_{ab}=1$. In general, there may be a proper subset $\Gamma_\iota\subset \Gamma$ with 
\[
\sum_{g\in \Gamma_\iota}\Pr[\gamma_{ab}=\g\mid \beta_{ab} = 1, \iota_{ab}=1]=1.
\]
This is an extreme version of the third bullet in Section \ref{sec:blockingmodel}. 

Indexing by disjunctions does not { necessarily} change the support; for example, we might index on the disjunction of agreement on the first 3 digits of the postal code and agreement of age within $\pm 5$ years, but include more stringent comparisons in $\gamma$ (such as matching on all postal code digits and ages within $\pm 1$ year). However, filtering restricts the support by design. Any changes in support should be explicitly reflected in subsequent modeling, which requires some modifications to the usual Fellegi-Sunter model. We discuss this further in Section \ref{sec:modeling}.

\subsection{Weights and Error Rates after Indexing}

Unless the various bias due to indexing is specifically addressed (as proposed in \cite{Jaro1989} for example) the estimated error rates are conditional on $\beta_{ab}=1$ after blocking (as well as $\iota_{ab}=1$ if blocking is followed by indexing by disjunctions or filtering). That is, under blocking we obtain estimates of
\begin{align}
\mu_\beta &= P(A_1\mid (a,b)\in U, \beta_{ab}=1) 
= \sum_{g\in\Gamma} P(A_1\mid  \gamma_{ab}= g, \beta_{ab}=1)\pi_{g\mid U, \beta}\\
\lambda_\beta &= P(A_3\mid (a,b)\in M, \beta_{ab}=1)
= \sum_{g\in\Gamma} P(A_3\mid  \gamma_{ab}= g, \beta_{ab}=1)\pi_{g\mid M, \beta}.
\end{align} 
After blocking and indexing by disjunctions or filtering we obtain
\begin{align}
\mu_{\beta,\iota} &= P(A_1\mid (a,b)\in U, \beta_{ab}=1, \iota_{ab}=1) 
= \sum_{g\in\Gamma} P(A_1\mid  \gamma_{ab}= g, \beta_{ab}=1, \iota_{ab}=1)\pi_{g\mid U, \beta, \iota}\\
\lambda_{\beta,\iota} &= P(A_3\mid (a,b)\in M, \beta_{ab}=1, \iota_{ab}=1) \\
&= \sum_{g\in\Gamma} P(A_3\mid  \gamma_{ab}= g, \beta_{ab}=1, \iota_{ab}=1)\Pr[\gamma_{ab}=g \mid (a,b)\in M, \beta_{ab}=1, \iota_{ab}=1]\\
&= \sum_{g\in\Gamma} P(A_3\mid  \gamma_{ab}= g, \beta_{ab}=1, \iota_{ab}=1)\pi_{g\mid M, \beta,\iota}.
\end{align} 
Based on the discussion above, for most decision rules we would expect estimates of $\lambda_\beta$ and $\lambda_{\beta, \iota}$ to be similar when indexing is functioning as intended. Both are conditional error rates and do not address error induced by indexing.

Since we expect $\pi_{g\mid U,\beta,\iota}>\pi_{g\mid U,\beta}$ for comparison vectors $g$ with many ones,  $\mu_{\beta, \iota}$ will tend to be much larger than $\mu_\beta$ for reasonable decision rules (which set $P(A_1\mid (a,b)\in U, \beta_{ab}=1, \iota_{ab}=1)$ or $P(A_1\mid (a,b)\in U, \beta_{ab}=1, \iota_{ab}=1)$ to zero for comparison vectors $g$ that are less likely to indicate matches). But higher rates are more tolerable after additional indexing or filtering -- the actual number of false matches is primarily of concern, and there are fewer total non-matching pairs under consideration. If there are $n_\beta$ non-matching pairs after blocking and $k_{\beta,\iota}$ of these are excluded in a subsequent indexing/filtering step then the expected number of false matches is $\mu_\beta n_\beta$ using blocking alone and $\mu_{\beta,\iota}(n_{\beta}-k_{\beta,\iota})$ using blocking and indexing/filtering. Setting 
\begin{equation}
\mu_{\beta, \iota} = \mu_\beta \frac{n_{\beta,}}{n_{\beta}-k_{\beta,\iota}}
\end{equation}
provides similar control of the total number of false matches under blocking alone and blocking with additional indexing/filtering. The unknown number of true non-matching pairs $n_\beta$ can be conservatively estimated by the total number of pairs remaining after blocking.

The effect of indexing by disjunctions or filtering on the weights is less obvious. Define
\begin{align}
w_{\g\mid \beta} &=  \frac{\Pr[\gamma_{ab}=\g\mid (a,b)\in M, \beta_{ab} = 1]
}{
	\Pr[\gamma_{ab}=\g\mid (a,b)\in U, \beta_{ab} = 1]
}\\ &= \frac{\pi_{\g\mid M, \beta}}{\pi_{\g\mid U, \beta}}\\
w_{\g\mid \beta, \iota} &= \frac{\Pr[\gamma_{ab}=\g\mid (a,b)\in M, \beta_{ab} = 1, \iota_{ab}=1]
}{
	\Pr[\gamma_{ab}=\g\mid (a,b)\in U, \beta_{ab} = 1, \iota_{ab}=1]
}\\ &= \frac{\pi_{\g\mid M, \beta, \iota}}{\pi_{\g\mid U, \beta, \iota}}.
\end{align}
For any comparison vector $g$ with $\Pr[\gamma_{ab}=\g\mid (a,b)\in U, \beta_{ab} = 1, \iota_{ab}=1]>0$,\footnote{Formally, we also require $\Pr[\iota_{ab}=1\mid (a,b)\in M, \beta_{ab}=1]$ and $\Pr[\iota_{ab}=1\mid (a,b)\in U, \beta_{ab}=1]$ to be nonzero. This will be the case under any practical indexing/filtering procedure.}

\begin{align}
w_{\g\mid  \beta, \iota}
= 
w_{\g\mid \beta}
\times
\frac{\Pr[\iota_{ab}=1 \mid \gamma_{ab}=\g, (a,b)\in M, \beta_{ab} = 1]
}{
	\Pr[\iota_{ab}=1 \mid \gamma_{ab}=\g, (a,b)\in U, \beta_{ab} = 1]
}
&\times 
\frac{\Pr[\iota_{ab}=1 \mid (a,b)\in U, \beta_{ab} = 1]
}{
	\Pr[\iota_{ab}=1 \mid (a,b)\in M, \beta_{ab} = 1]
}.\label{eq:weighteq}
\end{align}
When $\iota_{ab}$ is completely determined by $\gamma_{ab}$ this simplifies to

\begin{align}
w_{\g\mid \beta, \iota}
= 
w_{\g\mid \beta}
&\times 
\frac{\Pr[\iota_{ab}=1 \mid (a,b)\in U, \beta_{ab} = 1]
}{
	\Pr[\iota_{ab}=1 \mid (a,b)\in M, \beta_{ab} = 1]
}\label{eq:filterwt}
\end{align}
This condition will hold when indexing by disjunctions of elements in $\gamma$ (which includes filtering as a special case).  Since the second term of $\eqref{eq:filterwt}$ does not depend on $g$, in this case the rank order of $w_{g\mid \beta, \iota}$ agrees with the rank order of $w_{g\mid \beta}$.

In general, however, we have no such guarantee. If we assume that indexing/filtering is error-free (in that it does not exclude any truly matching pairs) we have the simple relationship

\begin{align}
w_{\g\mid  \beta, \iota}
&= 
w_{\g\mid \beta}
\times
\frac{\pi_{g\mid U, \beta}}{\pi_{g\mid U, \beta, \iota}}.
\label{eq:weighteqperfect}
\end{align}
The second term in \eqref{eq:weighteqperfect} will tend to vary across $g$, particularly when $\iota_{ab}$ is constructed from relaxed versions of some of the comparisons in $\gamma_{ab}$. This can alter the ranking that would be obtained from $w_{g\mid\beta}$ when using $w_{g\mid\beta,\iota}$ instead.

Similar calculations apply when comparing error rates and matching weights with and without blocking (before indexing by disjunctions/filtering). Overall it seems difficult to use the parameter estimates after indexing to make general statements about what the results would have been without indexing, even if we make generous assumptions about model specification and the errors induced by indexing. We prefer to focus explicitly on conditional versions of the parameters. When indexing by disjunctions or filtering this means our model must account for any changes in support, which requires extensions to models typically used in the Fellegi-Sunter framework.
%

\section{Modeling Record Pairs after Indexing by Disjunctions/Filtering}\label{sec:modeling}

Consider the contingency table formed by the binary comparison vectors. As noted above, after filtering or indexing by disjunctions the contingency table may be {\em incomplete} -- some cell counts are unobserved or fixed at zero \citep{10.2307/2528967,bishop1975discrete}. The number of incomplete cells can be large. For example, if we index by the disjunction of two out of $q$ total binary comparisons in $\gamma$, then $2^{(q-2)}$ of the cell counts in the table are unobserved after indexing. Subsequent filtering will generate more incomplete cells. 

Incomplete cells should be treated as either structural zeros or missing data. Treating the incomplete cells as missing effectively extrapolates from the pairs remaining after indexing by disjunctions/filtering to estimate the parameters in model \eqref{eq:fs-block0}-\eqref{eq:fs-block1}. In general, however, the estimates will be biased away from the estimates we would have gotten if we used blocking alone (data from the incomplete cells are not missing at random \citep{rubin1976inference}). On the other hand, treating the incomplete cells as structural zeros targets the parameters in \eqref{eq:fs-index0}-\eqref{eq:fs-index1} directly. The structural zero formulation is more appropriate for the following reasons:
\begin{enumerate}
\item In the structural zero formulation, the match/non-match probabilities, weights, error rates and decision rule thresholds 
are explicitly conditional on $\iota_{ab}=1$ and have support $\{\gamma: \gamma\in \Gamma_\iota\}$, the set of comparisons actually under consideration. This is in accordance with our discussion in the previous section and with \cite{FS}'s original recommendation to explicitly specify the comparison space.
\item 
The proportion of true matches after blocking, $p_{M\mid \beta}$, is typically much smaller than $p_{M\mid\beta,\iota }$ because the set of excluded pairs is composed disproportionately (or entirely) of non-matching records. From a parameter estimation perspective larger values for the proportion of matches are better (see e.g. \cite{winkler2006overview}, who suggests that at least 5\% of the record pairs under consideration should be matches for maximum likelihood estimates computed via EM to be reliable).
\item Under model misspecification the structural zero formulation may better approximate true values of relevant probabilities (and therefore error rates, decision rule thresholds, and matching weights).
Treating the incomplete cells as structural zeros and estimating the parameters of \eqref{eq:fs-index0}-\eqref{eq:fs-index1} by maximum likelihood yields the parameters that best approximate $P(\gamma_{ab} \mid \beta_{ab} = 1, \iota_{ab}=1)$ (in the Kullback-Leibler sense). In general these will be distinct from the parameters best approximating $P(\gamma_{ab} \mid \beta_{ab} = 1)$ or $P(\gamma_{ab})$, which are not of primary interest.
\end{enumerate}

In the saturated model accounting for the support restriction is trivial. But the saturated model in \eqref{eq:fs-index0}-\eqref{eq:fs-index1} will usually not be estimable. A natural extension of the conditional independence assumption \eqref{eq:ci} to models with structural zeros is a conditional {\em quasi-}independence model \citep{goodman1968analysis,fienberg1970quasi,bishop1975discrete}:

\begin{equation}
\tpi_{\g\mid M, \beta, \iota}\propto \prod_{j=1}^p \psi^{(j)}_{g(j)\mid M, \beta, \iota}\ind{g\in\Gamma_\iota},
\quad \tpi_{\g\mid U, \beta, \iota}\propto \prod_{j=1}^p \psi^{(j)}_{g(j)\mid U, \beta, \iota}\ind{g\in\Gamma_\iota}.\label{eq:qci}
\end{equation}
The $\psi$ parameters above are not identified without further constraints, but we are only concerned with the induced $m-$ and $u-$probabilities (which are identified).  

The conditional quasi-independence model is straightforward to estimate via EM. Let $\{\tpi^{(t)}_{\g\mid M, \beta, \iota}, \tpi^{(t)}_{\g\mid U, \beta, \iota} : g\in\Gamma_\iota \}$ and $p^{(t)}_{M\mid \beta, \iota}$ be the parameters at iteration $t$. The EM algorithm proceeds as follows:

\begin{itemize}
\item (E-Step) Compute the expected cell counts for matching and non-matching pairs:
\begin{align}
\tilde n_{g, M}^{(t+1)} &= 
n_g s_{g}^{(t)}\\
\tilde n_{g, U}^{(t+1)} &= 
n_g(1-s^{(t)}_{g}),
\end{align}
where $n_g$ is the number of record pairs with comparison vector $g$ and $s_{g}^{(t)}$ is the conditional probability that $(a,b)\in M$ given $\gamma_{ab}=g$ and the current values of the parameters:
\begin{equation}
s_{g}^{(t)} = \frac{p^{(t)}_{M\mid \beta, \iota}\pi^{(t)}_{\g\mid M, \beta,\iota} }
{p^{(t)}_{M\mid \beta, \iota}\pi^{(t)}_{\g\mid M, \beta,\iota} + (1-p^{(t)}_{M\mid \beta, \iota})\pi^{(t)}_{\g\mid U, \beta,\iota}}.
\end{equation}
\item (M-Step 1) Set 
\begin{align}
p^{(t+1)}_{M\mid \beta, \iota} = 
\frac{\sum_{g\in \Gamma_\iota}\tilde n^{(t+1)}_{g, M}}
{n}
\end{align}
\item (M-Step 2) Set
\begin{align}
\{\tpi^{(t+1)}_{\g\mid M, \beta, \iota} : g\in\Gamma_\iota \} = \argmax \sum_{g\in\Gamma_i} 
{\tilde n_{g, M}^{(t+1)}}\log\left(\tpi_{\g\mid M, \beta, \iota}\right)\\
\{\tpi^{(t+1)}_{\g\mid U, \beta, \iota} : g\in\Gamma_\iota \} = \argmax \sum_{g\in\Gamma_i} 
{\tilde n_{g, U}^{(t+1)}}\log\left(\tpi_{\g\mid U, \beta, \iota}\right),
\end{align}
where both maximizations are over the $|\Gamma_{\iota}|-$dimensional simplex. 
\end{itemize}
M-step 2 is the only step that deviates from the usual EM algorithm for the conditional independence model. The maximizations must be done numerically due to the support restrictions.

A simple approach is to use (quasi-)Poisson regression, recognizing that each maximization problem above can be recast as fitting a log-linear model under quasi-independence by maximum likelihood and employing the multinomial-Poisson transform \citep{baker1994}. The response vector includes {\em all} the cell counts, complete and incomplete, with zeros for the incomplete entries. The design matrix includes a main effect for each comparison as well as an indicator for each incomplete cell. The indicators force the estimates of incomplete cell probabilities to be zero. With a large number of cells alternative algorithms may be necessary, but this approach is feasible for binary comparisons and common values of $p$ (less than 11 or 12). R code implementing the EM algorithm appears in Appendix 1 and is posted online\footnote{\url{http://andrew.cmu.edu/~jsmurray/research/}}.

The conditional quasi-independence model can be extended along similar directions as the conditional independence model. For example, the $\psi$ parameters in \eqref{eq:qci} can be replaced by a log-linear model with interactions. However, modeling $P(\gamma_{ab}=g\mid \beta_{ab}=1, \iota_{ab}=1)$ directly may confer at least some degree of robustness to the conditional quasi-independence assumption, as we will see in the example below. 

\section{Example: Synthetic Data (RLdata10000)}\label{sec:example}

To illustrate the benefits of filtering and conditional quasi-independence models that account for it we compare the Fellegi-Sunter model under conditional quasi-independence using blocking and filtering to the standard Fellegi-Sunter model under conditional independence using blocking alone. We use a benchmark dataset (\texttt{RLdata10000}) distributed with the R package \texttt{RecordLinkage} \citep{recordlinkageR}. The dataset contains 9,000 distinct synthetic records of individuals. Each record has  names and dates of birth generated from real German population-level data. A random sample of 1,000 of the records were appended to the dataset and corrupted. The goal is to identify these duplicate records.

Details about the exact process used to corrupt the duplicated records are not available. However, simple statistical tests indicate that the conditional independence assumption does not hold. For example, after blocking the $\chi^2$ statistic for testing independence of agreement on first and last name among truly matching pairs is 162, with a numerically zero p-value. Therefore both models are misspecified.

The comparison vector comprises thresholded Jaro-Winkler scores for the comparisons on first and last name \citep{winkler90} and exact matching on day, month and year of birth. The Jaro-Winkler scores are thresholded at 0.9 here. The indexing scheme begins with a traditional blocking step retaining only pairs matching on first and last initial. For the conditional quasi-independence model this is followed by a filtering step which requires that records match on at least two of the five fields (first names, last name, and day, year or month of birth). This mimics the output of programs like BigMatch \citep{censusyancey02}. No true matching pairs are excluded in either step. The blocking step reduces the number of pairs under consideration from $(10,000\times 9,999)/2=49,995,000$ to $371,944$. After filtering, $34,896$ pairs remain. 


\subsection{Results}

\begin{figure}
 \centering
 \includegraphics[width=.95\textwidth]{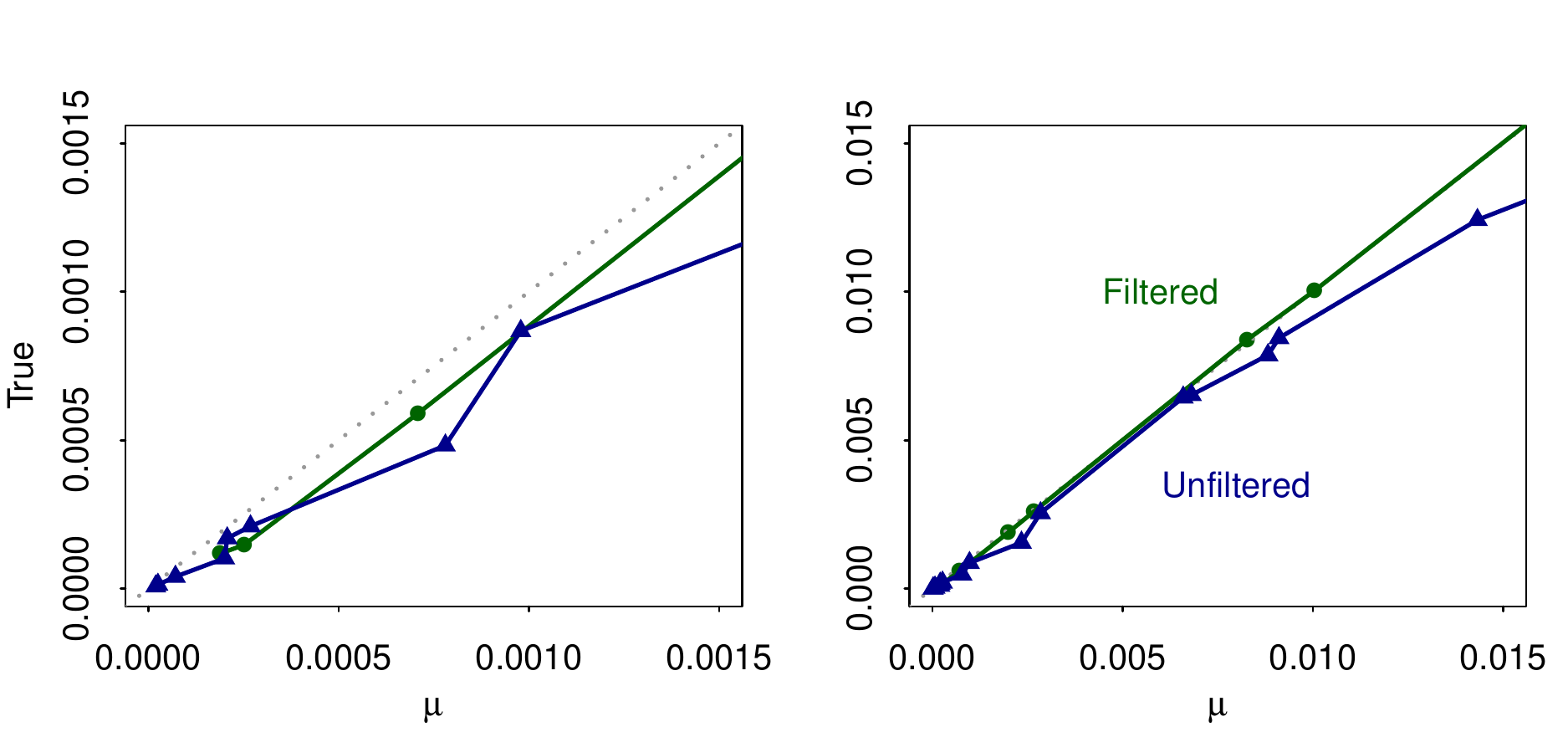}
  \includegraphics[width=.75\textwidth]{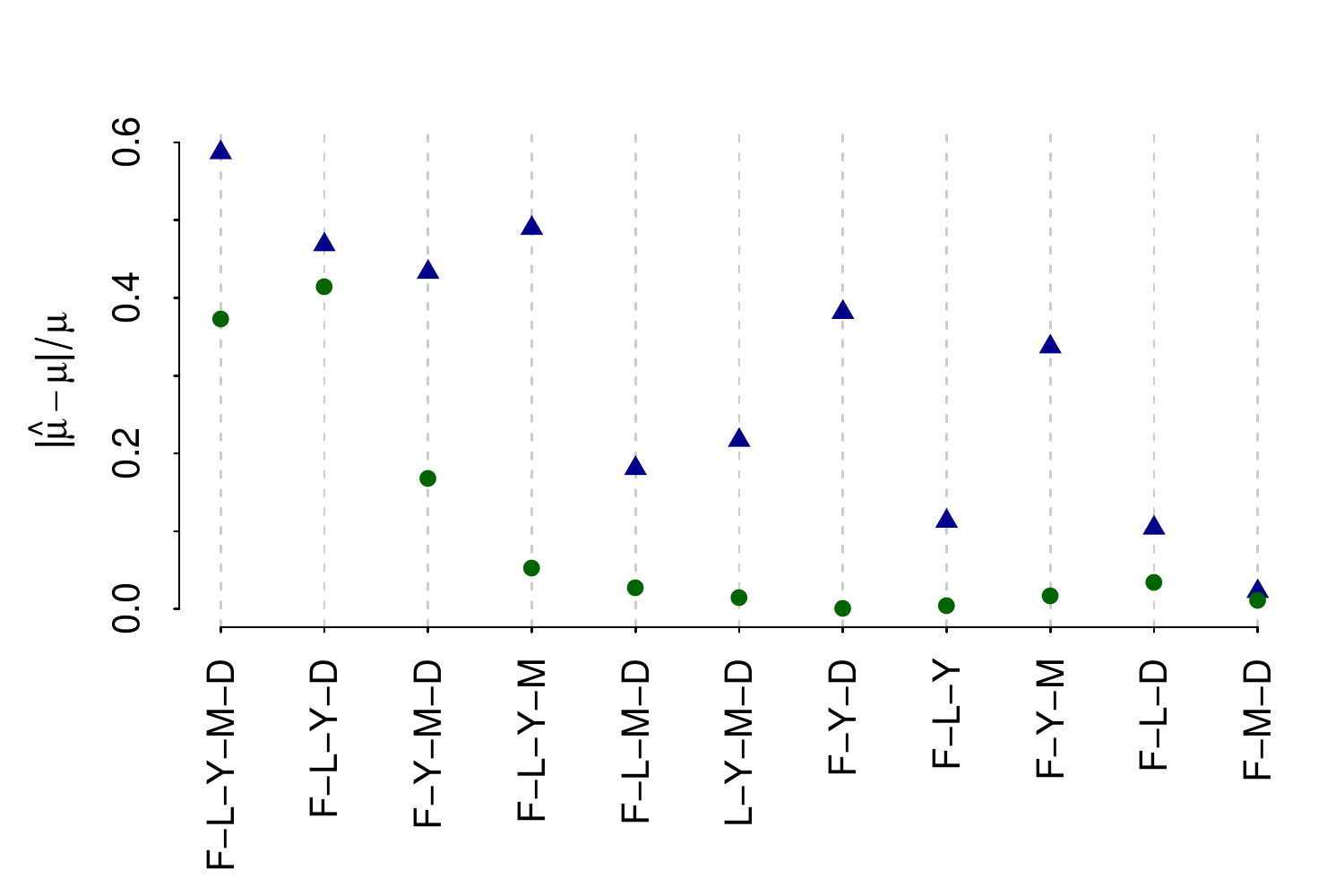}
 \caption{(Top) Estimated versus true values for the error rates $P(A_1\mid (a,b)\in U, \beta=1)$ (without filtering) and $P(A_1\mid (a,b)\in U, \beta=1, \iota=1)$ (with filtering) when the match threshold $T_\mu$ is chosen to have nominal error rate $\mu$. Points indicate values of $\mu$ for which the $T_\mu$ changes; intermediate values of $\mu$ rely on a randomized decision rule. (Bottom) Absolute relative discrepancy in the estimate of$P(A_1\mid (a,b)\in U, \beta=1)$ or $P(A_1\mid (a,b)\in U, \beta=1, \iota=1)$ as comparison patterns are added to the set of declared matches. Here comparison patterns are denoted by the fields of agreement (first name, last name, year, month and day of birth), and the threshold used for string comparisons is 0.9}
 \label{fig:rates} 
\end{figure}

The estimates of match proportions are $\hat p_{M\mid \beta} = .0029$ and $\hat p_{M\mid \beta, \iota} = 0.032$. Both are reasonable, since the true number of matching pairs is $1,000$. The weights from the filtered and unfiltered models give the same rank order over $\Gamma_\iota$. Figure \ref{fig:rates} shows that using filtering and a conditional quasi-independence model gives improved estimates of error rates. The error rates themselves are not directly comparable, as noted in Section \ref{sec:modeling}, but the filtered error rates are more relevant and better calibrated overall. For a more comparable measure we consider the relative discrepancy between nominal and actual error rates as comparison vectors are successively added to the match region of the decision rule. The bottom panel of Figure \ref{fig:rates} shows that the relative discrepancy is uniformly better under filtering and the conditional quasi-independence model.

\begin{figure}
 \centering
 \includegraphics[width=.75\textwidth]{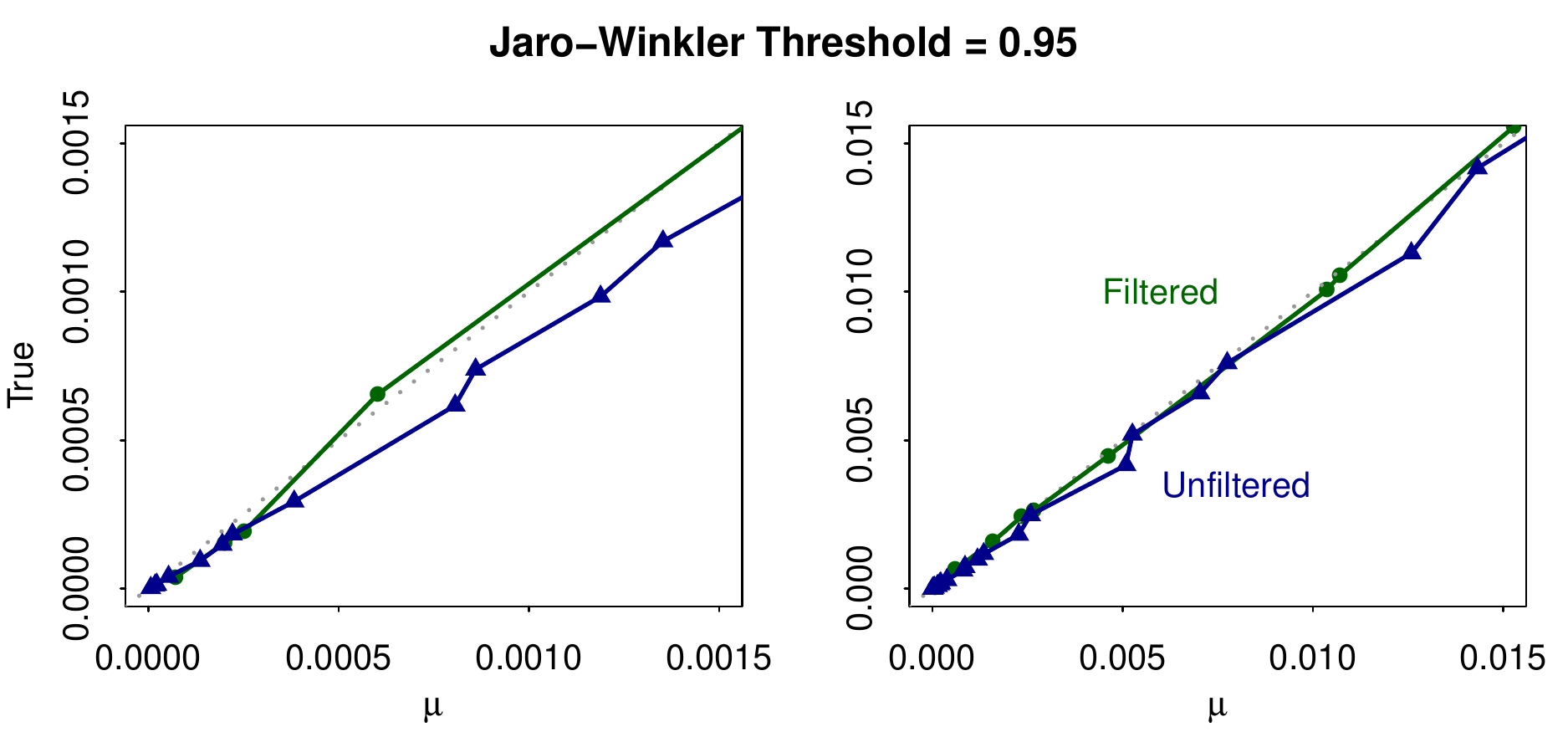}
  \includegraphics[width=.75\textwidth]{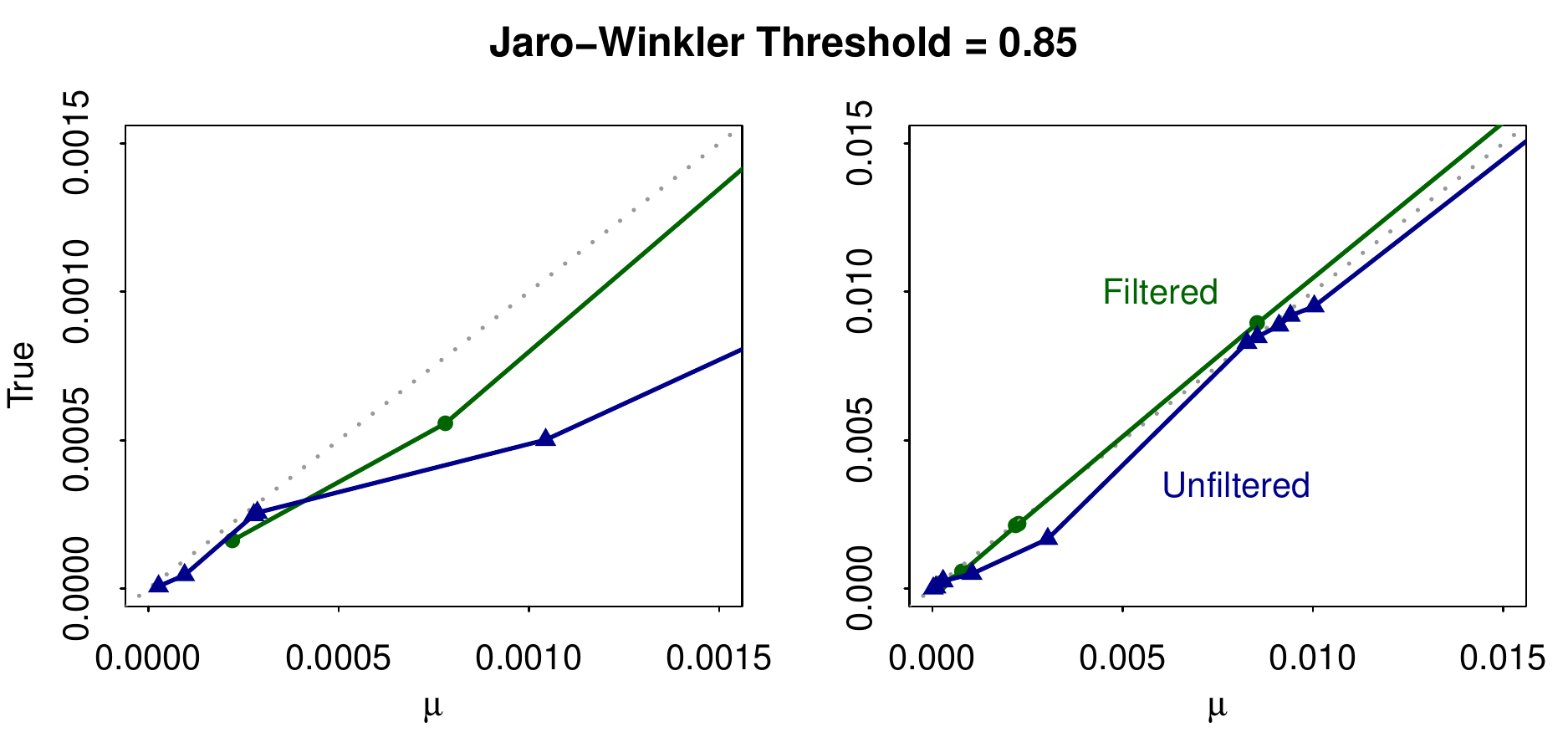}
   \includegraphics[width=.75\textwidth]{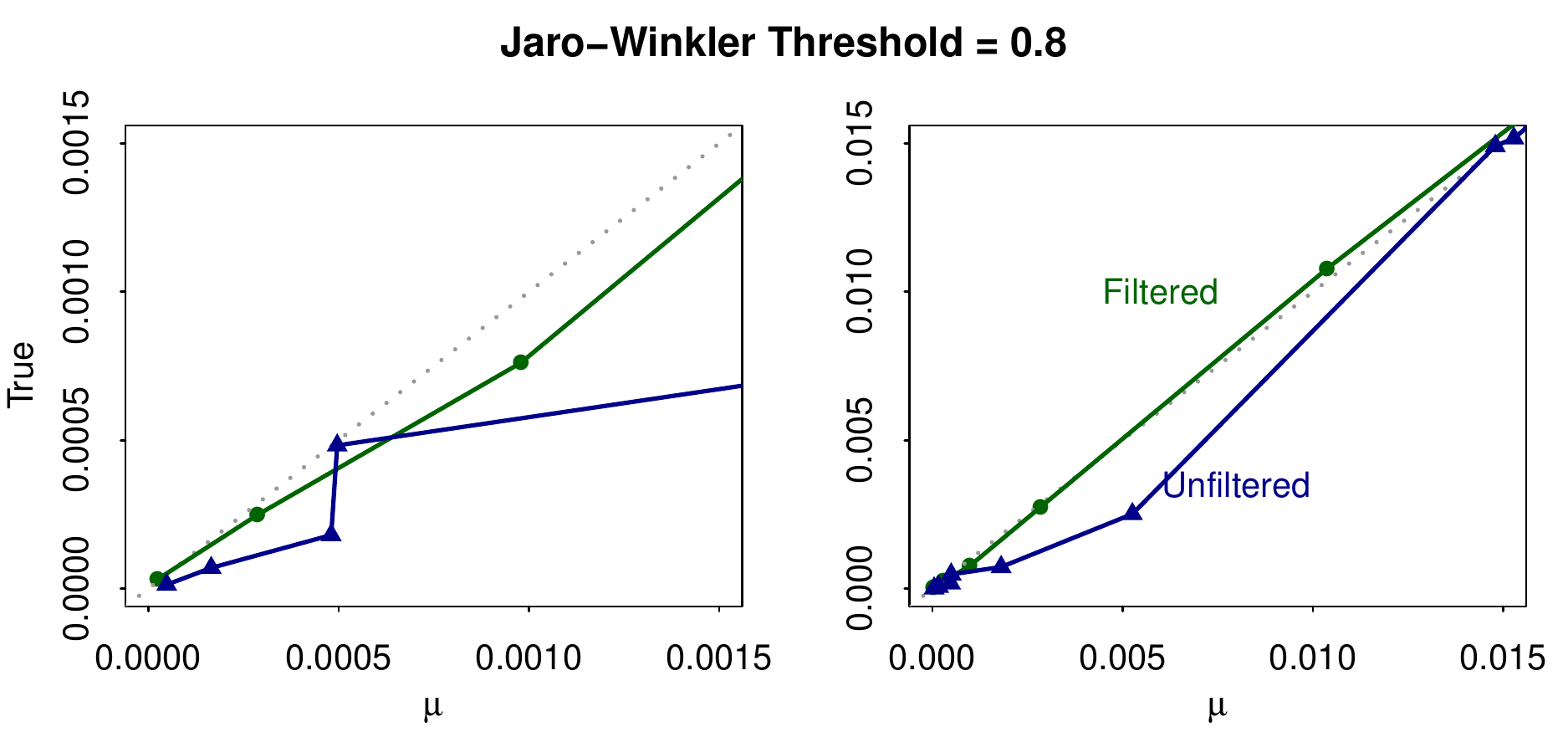}
 \caption{Estimated versus true values for the error rates $P(A_1\mid (a,b)\in U, \beta=1)$ (without filtering) and $P(A_1\mid (a,b)\in U, \beta=1, \iota=1)$ (with filtering) when the match threshold $T_\mu$ is chosen to have nominal error rate $\mu$. Points indicate values of $\mu$ for which the $T_\mu$ changes; intermediate values of $\mu$ rely on a randomized decision rule. Each pair of plots corresponds to a different choice of threshold for the Jaro-Winkler string comparison scores.}
 \label{fig:ratesthresh} 
\end{figure}

We tried a variety of other thresholds for the Jaro-Winkler scores. 
Error rate curves appear in Figure \ref{fig:ratesthresh}. Again, the filtered error rate estimates are better calibrated. Across different thresholds the cells with highest weight typically had the same rank order with and without filtering. However, for some threshold values the rank order of cells with intermediate weights varied, so reproducing the relative discrepancy plots in the bottom panel of Figure \ref{fig:rates} was not possible. But for the cells with highest weight the relative discrepancy was lower with filtering than using blocking alone.

\section{Conclusion}\label{sec:conclusion}

We have described the effects of indexing, blocking and filtering on subsequent inference of record linkage structure within the Fellegi-Sunter framework.  Explicitly modeling the effects of indexing, and especially filtering, clarifies the interpretation of error rates and enhances their estimation (which also improves decision rules). The effects of filtering in particular will be the greatest when the files lack a small number of highly discriminative fields for use in indexing.  This situation seems to be common in practice. Some of the impacts of indexing have been discussed in the literature, but modern applications of record linkage index, block or filter without making subsequent adjustments to the model for record pairs.

In related work \cite{winkler2006automatically} considered fixing a subset of highly likely/unlikely matching pairs as sure matches/non-matches during parameter estimation. This is closely related to filtering, which declares highly unlikely matching pairs as sure non-matches, but ignores them during parameter estimation. Blending the two strategies could prove fruitful. The most extreme comparison vectors could be filtered and ignored in parameter estimation, while less extreme but still very unlikely values could be fixed as sure non-matches to aid in parameter estimation. 

We have focused on three simple but extremely common indexing methods: blocking, indexing by disjunctions, and filtering. A range of other techniques for indexing exist, including sophisticated approaches based on various hashing algorithms \citep{christen2012indexing,steorts2014comparison}. These are perhaps best understood as fast approximations to filtering, and our discussion here applies more or less directly (especially if these indexing methods are followed by a subsequent filtering step to remove unlikely pairs that escape indexing).

The developments in this paper have applicability outside the traditional Fellegi-Sunter framework. With some relatively straightforward modifications the conditional quasi-independence model (or generalizations thereof) can be applied within in \cite{sadinle2013generalized}'s multiple-file generalization of the Fellegi-Sunter framework. Interestingly, \cite{sadinle2013generalized} include an example showing that blocking yields better estimates of error rates even when it is computationally feasible to make all the comparisons. But in that setting filtering is a better choice than blocking, since filtering does not require high-quality blocking keys and will only remove pairs that are {\em known} to have comparison vectors which are unlikely to indicate a match. Our example above shows that filtering can accrue similar benefits in error rate estimation. 

The conditional quasi-independence model introduced here is applicable within Bayesian approaches that rely on comparison vector-based likelihoods (e.g. \cite{mcglincy2004bayesian,larsen2005hierarchical,larsen2012experiment,sadinle2014detecting,sadinle2016bipartite}). Indexing, blocking and filtering all reduce the space of possible linkage structures that must be traversed during Markov Chain Monte Carlo, and may prove indispensable in scaling these methods to larger datasets. 
A challenge in this context is efficiently sampling the parameters determining the $m-$ and $u-$probabilities. The data augmentation algorithm introduced in \cite{manrique2014bayesian} is not immediately applicable, but could possibly be adapted to this purpose.

The implications of various forms of indexing for Bayesian methods that utilize full probability models for the raw data rather than comparisons are less clear (e.g. \cite{Tancredi2011,gutman2013bayesian,steorts2013bayesian,steorts2015entity}). Most of these either use do not indexing or rely on blocking, but filtering can significantly reduce the space of possible linkage structures and may play more of a role as these methods are scaled to larger problems. Indexing and filtering can complicate elicitation of joint probability models for the fields in each file. For example, when the two files have limited overlap large and non-random subsets of records may be excluded from consideration entirely. The retained records are not exchangeable with the excluded records and prior beliefs about the complete file will not immediately transfer to the retained records.

The implications for supervised record linkage, which utilizes a set of known matching and non-matching pairs to predict unlabeled pairs, are also unclear. The biased sampling due to indexing and filtering may be largely irrelevant since the unlabeled record pairs come from a similarly biased sample. However, indexing concentrates the predictors on a subspace (e.g., $\Gamma_\iota$ in the context of this paper). Perhaps this dimension-reducing effect could be exploited to enhance prediction; \cite{Ventura2015}'s blend of random forests and hierarchical clustering involves some similar ideas in this direction.

\bibliographystyle{apalike}
\bibliography{bib}

\begin{thebibliography}{}

\bibitem[Baker, 1994]{baker1994}
Baker, S.~G. (1994).
\newblock The multinomial-{P}oisson transformation.
\newblock {\em Journal of the Royal Statistical Society. Series D (The
  Statistician)}, 43(4):495--504.

\bibitem[Baxter et~al., 2003]{baxter2003comparison}
Baxter, R., Christen, P., and Churches, T. (2003).
\newblock A comparison of fast blocking methods for record linkage.
\newblock In {\em ACM SIGKDD}, volume~3, pages 25--27.

\bibitem[Bishop et~al., 1975]{bishop1975discrete}
Bishop, Y.~M., Fienberg, S.~E., and Holland, P.~W. (1975).
\newblock {\em Discrete multivariate analysis: theory and practice}.
\newblock Springer Science \& Business Media.

\bibitem[Borg and Sariyar, 2015]{recordlinkageR}
Borg, A. and Sariyar, M. (2015).
\newblock {\em RecordLinkage: Record Linkage in R}.
\newblock R package version 0.4-8.

\bibitem[Christen, 2012a]{Christen201207}
Christen, P. (2012a).
\newblock {\em Data Matching: Concepts and Techniques for Record Linkage,
  Entity Resolution, and Duplicate Detection (Data-Centric Systems and
  Applications)}.
\newblock Springer, 2012 edition.

\bibitem[Christen, 2012b]{christen2012indexing}
Christen, P. (2012b).
\newblock A survey of indexing techniques for scalable record linkage and
  deduplication.
\newblock {\em Knowledge and Data Engineering, IEEE Transactions on},
  24(9):1537--1555.

\bibitem[Fellegi and Sunter, 1969]{FS}
Fellegi, I.~P. and Sunter, A.~B. (1969).
\newblock A theory for record linkage.
\newblock {\em Journal of the American Statistical Association},
  64(328):1183--1210.

\bibitem[Fienberg, 1970]{fienberg1970quasi}
Fienberg, S.~E. (1970).
\newblock Quasi-independence and maximum likelihood estimation in incomplete
  contingency tables.
\newblock {\em Journal of the American Statistical Association},
  65(332):1610--1616.

\bibitem[Fienberg, 1972]{10.2307/2528967}
Fienberg, S.~E. (1972).
\newblock The analysis of incomplete multi-way contingency tables.
\newblock {\em Biometrics}, 28(1):177--202.

\bibitem[Goodman, 1968]{goodman1968analysis}
Goodman, L.~A. (1968).
\newblock The analysis of cross-classified data: Independence,
  quasi-independence, and interactions in contingency tables with or without
  missing entries: {R}.{A}. {F}isher memorial lecture.
\newblock {\em Journal of the American Statistical Association},
  63(324):1091--1131.

\bibitem[Gutman et~al., 2013]{gutman2013bayesian}
Gutman, R., Afendulis, C.~C., and Zaslavsky, A.~M. (2013).
\newblock A {B}ayesian procedure for file linking to analyze end-of-life
  medical costs.
\newblock {\em Journal of the American Statistical Association},
  108(501):34--47.

\bibitem[Herzog et~al., 2007]{HerzogScheurenWinkler200705}
Herzog, T.~N., Scheuren, F.~J., and Winkler, W.~E. (2007).
\newblock {\em Data Quality and Record Linkage Techniques}.
\newblock Springer, 2007 edition.

\bibitem[Jaro, 1989]{Jaro1989}
Jaro, M.~a. (1989).
\newblock {Advances in Record-Linkage Methodology as Applied to Matching the
  1985 Census of Tampa, Florida}.
\newblock {\em Journal of the American Statistical Association},
  84(406):414--420.

\bibitem[Larsen, 2005]{larsen2005hierarchical}
Larsen, M.~D. (2005).
\newblock Advances in record linkage theory: Hierarchical {B}ayesian record
  linkage theory.
\newblock In {\em Proceedings of the Section on Survey Research Methods}.

\bibitem[Larsen, 2012]{larsen2012experiment}
Larsen, M.~D. (2012).
\newblock An experiment with hierarchical {B}ayesian record linkage.
\newblock {\em arXiv preprint arXiv:1212.5203}.

\bibitem[Manrique-Vallier and Reiter, 2014]{manrique2014bayesian}
Manrique-Vallier, D. and Reiter, J.~P. (2014).
\newblock Bayesian estimation of discrete multivariate latent structure models
  with structural zeros.
\newblock {\em Journal of Computational and Graphical Statistics},
  23(4):1061--1079.

\bibitem[McGlincy, 2004]{mcglincy2004bayesian}
McGlincy, M.~H. (2004).
\newblock A {B}ayesian record linkage methodology for multiple imputation of
  missing links.
\newblock In {\em Proceedings of the Section on Survey Research Methods
  4001–4008. Amer. Statist.Assoc.}

\bibitem[Newcombe and Kennedy, 1962]{newcombe1962record}
Newcombe, H.~B. and Kennedy, J.~M. (1962).
\newblock Record linkage: Making maximum use of the discriminating power of
  identifying information.
\newblock {\em Communications of the ACM}, 5(11):563--566.

\bibitem[Newcombe et~al., 1959]{Newcombe1959}
Newcombe, H.~B., Kennedy, J.~M., Axford, S., and James, A.~P. (1959).
\newblock Automatic linkage of vital records computers can be used to extract
  ``follow-up" statistics of families from files of routine records.
\newblock {\em Science}, 130(3381):954--959.

\bibitem[Rubin, 1976]{rubin1976inference}
Rubin, D.~B. (1976).
\newblock Inference and missing data.
\newblock {\em Biometrika}, 63(3):581--592.

\bibitem[Sadinle, 2014]{sadinle2014detecting}
Sadinle, M. (2014).
\newblock Detecting duplicates in a homicide registry using a bayesian
  partitioning approach.
\newblock {\em The Annals of Applied Statistics}, 8(4):2404--2434.

\bibitem[Sadinle, 2016]{sadinle2016bipartite}
Sadinle, M. (2016).
\newblock Bayesian estimation of bipartite matchings for record linkage.
\newblock {\em Journal of the American Statistical Association (to appear)}.

\bibitem[Sadinle and Fienberg, 2013]{sadinle2013generalized}
Sadinle, M. and Fienberg, S.~E. (2013).
\newblock A generalized {F}ellegi--{S}unter framework for multiple record
  linkage with application to homicide record systems.
\newblock {\em Journal of the American Statistical Association},
  108(502):385--397.

\bibitem[Sadosky et~al., 2015]{sadosky2015blocking}
Sadosky, P., Shrivastava, A., Price, M., and Steorts, R.~C. (2015).
\newblock Blocking methods applied to casualty records from the syrian
  conflict.
\newblock {\em arXiv preprint arXiv:1510.07714}.

\bibitem[Steorts, 2015]{steorts2015entity}
Steorts, R.~C. (2015).
\newblock Entity resolution with empirically motivated priors.
\newblock {\em Bayesian Analysis}, 10(4):849--875.

\bibitem[Steorts et~al., 2016]{steorts2013bayesian}
Steorts, R.~C., Hall, R., and Fienberg, S.~E. (2016).
\newblock A {B}ayesian approach to graphical record linkage and de-duplication.
\newblock {\em Journal of the American Statistical Association (to appear)}.

\bibitem[Steorts et~al., 2014]{steorts2014comparison}
Steorts, R.~C., Ventura, S.~L., Sadinle, M., and Fienberg, S.~E. (2014).
\newblock A comparison of blocking methods for record linkage.
\newblock In {\em Privacy in Statistical Databases}, pages 253--268. Springer.

\bibitem[Tancredi and Liseo, 2011]{Tancredi2011}
Tancredi, A. and Liseo, B. (2011).
\newblock {A hierarchical {B}ayesian approach to record linkage and population
  size problems}.
\newblock {\em Annals of Applied Statistics}, 5(2 B):1553--1585.

\bibitem[Thibaudeau, 1993]{thibaudeau1993discrimination}
Thibaudeau, Y. (1993).
\newblock The discrimination power of dependency structures in record linkage.
\newblock {\em Survey Methodology}, 19:31--38.

\bibitem[Ventura, 2015]{Ventura2015}
Ventura, S. (2015).
\newblock {\em Large-Scale Classification and Clustering Methods with
  Applications in Record Linkage}.
\newblock PhD thesis, Carnegie Mellon University.

\bibitem[Winkler et~al., 2010]{winkler2010fast}
Winkler, W., Yancey, W., and Porter, E. (2010).
\newblock Fast record linkage of very large files in support of decennial and
  administrative records projects.
\newblock In {\em Proceedings of the Section on Survey Research Methods,
  American Statistical Association}.

\bibitem[Winkler, 1988]{winkler1988using}
Winkler, W.~E. (1988).
\newblock Using the {E}{M} algorithm for weight computation in the
  {F}ellegi-{S}unter model of record linkage.
\newblock In {\em Proceedings of the Section on Survey Research Methods,
  American Statistical Association}, volume 667, page 671.

\bibitem[Winkler, 1990]{winkler90}
Winkler, W.~E. (1990).
\newblock String comparator metrics and enhanced decision rules in the
  fellegi-sunter model of record linkage.
\newblock In {\em Proceedings of the Section on Survey Research Methods}, pages
  354--359.

\bibitem[Winkler, 1993]{winkler1993improved}
Winkler, W.~E. (1993).
\newblock Improved decision rules in the fellegi-sunter model of record
  linkage.
\newblock In {\em ASA Proceedings of Survey Research Methods Section}.

\bibitem[Winkler, 2006a]{winkler2006automatically}
Winkler, W.~E. (2006a).
\newblock Automatically estimating record linkage false match rates.
\newblock In {\em ASA Proceedings of Survey Research Methods Section}.

\bibitem[Winkler, 2006b]{winkler2006overview}
Winkler, W.~E. (2006b).
\newblock Overview of record linkage and current research directions.
\newblock Technical Report Statistical Research Report Series RRC2006/02, U.S.
  Bureau of the Census, Washington, D.C.

\bibitem[Yancey, 2002]{censusyancey02}
Yancey, W.~E. (2002).
\newblock Big{M}atch: {A} program for extracting probable matches from a large
  file for record linkage.
\newblock Technical Report Statistical Research Report Series RRC2002/01, U.S.
  Bureau of the Census, Washington, D.C.

\end{thebibliography}

\appendix
\section{Appendix: Example R Code}

The code below is also available from \url{http://andrew.cmu.edu/~jsmurray/research/}.

\begin{lstlisting}[language=R]
library(RecordLinkage)

# Jaro-Winkler cutoff
cutoff = 0.9

######################################################################
# Load and process the data
######################################################################
data("RLdata10000")
dat = RLdata10000

dat$fi = substr(dat$fname_c1, 1, 1)
dat$li = substr(dat$lname_c1, 1, 1)
dedup = compare.dedup(dat, blockfld=c(8,9), exclude = c(2,4,8,9), 
                      strcmp = c(1,3), 
                      identity=identity.RLdata10000)

pairs = dedup$pairs[,-c(1,2)]
pairs$fname_c1 = as.numeric(pairs$fname_c1>=cutoff)
pairs$lname_c1 = as.numeric(pairs$lname_c1>=cutoff)

tdf = as.data.frame(table(pairs[,-ncol(pairs)]))

keep = rowSums(sapply(tdf[,-ncol(tdf)], as.numeric)-1)>=2
  
trunc_tdf = tdf
trunc_tdf[!keep,ncol(tdf)] = 0
counts = trunc_tdf$Freq
n = sum(counts)

######################################################################
# Build the design matrix
######################################################################
main.eff = matrix(as.numeric(as.matrix(tdf[,1:5])), nrow=nrow(tdf))
getind = function(s, n) {rr = rep(0, n); rr[s]=1; rr }
zero.indicator = sapply(which(counts==0), getind, n=nrow(trunc_tdf))
des = data.frame(main.eff, I=zero.indicator)

######################################################################
# Control settings for the EM algorithm
######################################################################
maxiter = 1000
tol = 1e-6 # Stopping criterion

######################################################################
# Begin EM algorithm
######################################################################

# Set initial values

# p_{M \mid beta, iota}
pM = 0.1

# pi_{g\mid U, beta, iota}
# Approximately the observed frequencies, since the
# total number of matches is small (add 2 to cell counts
# to avoid issues from sampling zeros)
piU = (counts + 2)/sum(counts + 2)
piU = piU*as.numeric(keep)
piU = piU/sum(piU)

# pi_{g\mid M, beta, iota}
# Truncated conditional independence model with 
# P(agree | match) = 0.95
piM = 0.95^rowSums(main.eff)*0.05^(5- rowSums(main.eff))
piM = piM*as.numeric(keep)
piM = piM/sum(piM)

for(i in 1:maxiter) {
  # E step
  cprobM = (1-pM)*piU/((1-pM)*piU + pM*piM)
  nU = counts*cprobM
  nM = counts*(1-cprob0M)
  nU[!keep] = 0
  nM[!keep] = 0
  
  # M step
  glm.fit.0 = glm(y~., data=data.frame(y=nU, des),
                  family="quasipoisson")
  glm.fit.1 = glm(y~., data=data.frame(y=nM, des),
                  family="quasipoisson")
  pM = sum(nM)/n
  
  piU.old = piU
  piM.old = piM
  
  g = function(fit, keep) {
    logwt = predict(fit)
    logwt = logwt - max(logwt)
    wt = as.numeric(keep)*exp(logwt)
    wt/sum(wt)
  }
  piU = g(glm.fit.0, keep)
  piM = g(glm.fit.1, keep)
  
  if (max(abs(log(piM/piU)[keep] - log(piM.old/piU.old)[keep])) < tol) break
}
\end{lstlisting}
\end{document}